\documentclass[sigconf]{acmart}

\AtBeginDocument{%
  \providecommand\BibTeX{{%
    \normalfont B\kern-0.5em{\scshape i\kern-0.25em b}\kern-0.8em\TeX}}}

\copyrightyear{2022} 
\acmYear{2022} 
\setcopyright{acmlicensed}\acmConference[CHI '22 Extended Abstracts]{CHI Conference on Human Factors in Computing Systems Extended Abstracts}{April 29-May 5, 2022}{New Orleans, LA, USA}

\acmBooktitle{CHI Conference on Human Factors in Computing Systems Extended Abstracts (CHI '22 Extended Abstracts), April 29-May 5, 2022, New Orleans, LA, USA}
\acmPrice{15.00}
\acmDOI{10.1145/3491101.3519703}
\acmISBN{978-1-4503-9156-6/22/04}

\begin{document}

\title{Undoing Seamlessness: Exploring Seams for Critical Visualization}

\author{Nicole Hengesbach}
\email{N.Hengesbach@warwick.ac.uk}
\orcid{0000-0001-7918-6343}
\affiliation{%
  \institution{University of Warwick, United Kingdom}
}


\begin{abstract}
 While seamful design has been part of discourses and work within HCI contexts for some time, it has not yet been fully explored in data visualization design. At the same time, critics of visualization have been arguing that the representation of data as contextual, contingent, relational, partial, heterogeneous, and situated is currently lacking in visualization. Seamful visualization promises a fresh perspective on visualization design as we seek to find more expressive encodings and novel approaches to representing data that acknowledge their wider qualities and limitations. By consulting seams in other realms and exploring existing seams and seamfulness in visualization, this paper offers a foundation for conceptualizing seamful visualization, points towards the value of seams and seamfulness in critical visualization, and proposes principles for engaging with seamful visualization in practice and research.
\end{abstract}



\maketitle

\section{Introduction}
Data visualizations often display data in seamless interfaces with a clean and usually aesthetically pleasing design and seamlessly implemented features, like zoom, or transitions between views. This aspiration to seamlessness is common in data practices, favoring clarity and cleanliness while data are seen as something that “need to be tidied and tamed” \cite{dignazio_data_2020}. In critical data visualization, this paper argues, we do not want this cleanliness as it distorts, hides, and disguises the complex reality of data, data practices, and phenomena from the real-world.

Through a clean and clear layout, seamlessness in visualizations may afford usability, readability, effectiveness, and the neat integration of tasks and features. However, this seamlessness doesn’t necessarily encourage critical engagements with data: it fails to represent the wider qualities of the data–phenomenon relationship \cite{offenhuber_data_2019}, that is how data may be contextual, contingent, relational, partial, heterogeneous and situated \cite{loukissas_all_2019, kitchin_towards_2014}, and the limitations of data, e.g., how they may be incorrect, inaccurate, inconsistent, or incomplete. Considering, surfacing, and representing such qualities and limitations may be of particular interest in terms of visualizations that are publicly available, widely shared, and of public interest, created and published by newspapers, citizen groups, or public institutions.

To encourage critical engagements with data in such contexts, we could specifically highlight the limitations and qualities of data, but while we have concepts and methods to critically study and analyze data sets, settings \cite{bates_data_2016, dignazio_creative_2017, loukissas_taking_2017, loukissas_place_2016, loukissas_all_2019, neff_critique_2017, van_es_towards_2017}, and data practices \cite{concannon_applying_2018, gabrys_just_2016, kitchin_towards_2014, pritchard_re-calibrating_2018, reed_discovering_2017, Richterich_2018}, it remains challenging to include cues about such wider qualitative aspects of data. Approaching critical visualization as the opposite to the prevalent seamlessness leads us to explore seamful visualization. This promises to provide stepping-stones for a visual language and design principles that expand our ability to represent the limitations and qualities of data and aspects of the data–phenomenon relationship.

This work is motivated by previous work on critical visualization \cite{dignazio_feminist_2016, offenhuber_data_2019, dork_critical_2013}, (critical) data studies \cite{boyd_critical_2012, kitchin_towards_2014, kitchin_data_2021, loukissas_all_2019}, and data feminism \cite{dignazio_data_2020}, and continues the seamful design discussion, established in ubiquitous computing and human-computer interaction (HCI) \cite{inman_beautiful_2019}. By tracing already existing occurrences of seamfulness and seams in visualization, this paper seeks to rekindle and expand the discussion on seamful visualization, explore its potential for critical visualization practice, and offer starting points for a conceptualization of seamful visualization.

\section{Background}
According to {\itshape Merriam-Webster} \cite{merriam-webster_definition_nodate} ‘seamless’ describes something as “having no awkward transitions, interruptions, or indications of disparity” or as “moving from one thing to another easily and without any interruptions or problems.” Synonyms include ‘flawless’, ‘immaculate’, ‘perfect’, ‘unblemished’. In the literature, the difference between ‘seamful’ and ‘seamless’ is summarized by Inman \& Ribes \cite{inman_beautiful_2019} as follows: “Roughly, seamless design emphasizes clarity, simplicity, ease of use, and consistency to facilitate technological interaction. In contrast, seamful design emphasizes configurability, user appropriation, and revelation of complexity, ambiguity, or inconsistency.” 

\subsection{Seamful design}
The discourse on seamful and seamless design was first initiated with Weiser's work on seamlessness in ubiquitous computing \cite{weiser_computer_1999, weiser_creating_1994, weiser_ubiquitous_1994} and his “[call] for well-designed configurability and strategic revelation of complexity, error, or backgrounded tasks.” \cite{inman_beautiful_2019} Following this, seamful design has been explored widely within HCI and ubiquitous computing, for instance, by Chalmers and colleagues who argue for seamfulness \cite{chalmers_seamful_2004, chalmers_seamful_2003} and explore seamful design in contexts of different ubiquitous computing systems, such as wearable computing \cite{chalmers_seamful_2003-2}, (social) navigation and mobile computing \cite{chalmers_social_2004}, and games \cite{chalmers_gaming_2005, bell_interweaving_2006}.

Inman \& Ribes \cite{inman_beautiful_2019} track this discourse, situate the seamful\slash seamless discussion within the context of the “strategic revelation and concealment of human and technological operations,” and find central themes in seamful design: in\slash visibility, uncertainty, appropriation, and time and interaction histories. Early work in HCI highlights seamful design as an opportunity to address the seams between different systems, or between the virtual and the real world, and more recent work acknowledges the critical potential of seamfulness. As such, recent work explores seamful design to surface algorithmic processes \cite{eslami_first_2016}, transformations in data \cite{inman_data_2018}, to make explainable AI more trustworthy \cite{ehsan_explainability_2021}, and to design a networked media artwork that combines seamfulness with openness and slowness \cite{bachler_slowness_2020}.

\subsection{Seamfulness outside of HCI}
Seamful design and, more so, seamfulness as a conceptual lens have been borrowed from HCI to be discussed in terms of learning in professional education \cite{fawns_seamful_2021} and collaboration in research environments \cite{mauro_towards_2017}. Seams as a concept have been playing a central role in recent infrastructure studies that started with Vertesi’s work on seamful spaces \cite{vertesi_seamful_2014}. Seamlessness and seamfulness have also been part of discourses considering smart cities \cite{dignazio_seamful_2019, raetzsch_weaving_2019, swist_assemblages_2019, zandbergen_unfinished_2020}. For instance, D’Ignazio et al. \cite{dignazio_seamful_2019} argue to make visible data collection and usage through seamful urban interfaces—urban interfaces being “the ways in which people are invited to interact at all stages with a city’s digital infrastructure during its ideation, installation and deployment”—that encourage seamful interaction, supporting their key argument that “[for] cities to be equitable and inclusive, these interfaces need to be visible.”

\subsection{Seamful visualization}
Previously, seamful visualization and seams in visualization have only been considered infrequently. Though previous work hints at the usefulness and potential of seamful visualization, it has not yet been sufficiently explored. In technical terms, seams in visualization have been explored in terms of dealing with seams between multiple monitors \cite{mackinlay_wideband_2004, ni_survey_2006}, and Chalmers \cite{chalmers_seamful_2003-1} presents a ‘seamful map’ that displays good\slash better and bad\slash worse signal strength of network service by plotting accordingly colored tiles onto an aerial photograph.

Hinting at the potential of seamful visualization for representing limitations of data, Sengers \& Gaver \cite{sengers_staying_2006} argue that seamful design can enable multiple interpretations and underline that “seamful designs explicitly represent the limitations and uncertainties in data, allowing users to make up their own minds about how to interpret
\begin{figure}[t]
  \centering
  \includegraphics[width=0.9\linewidth]{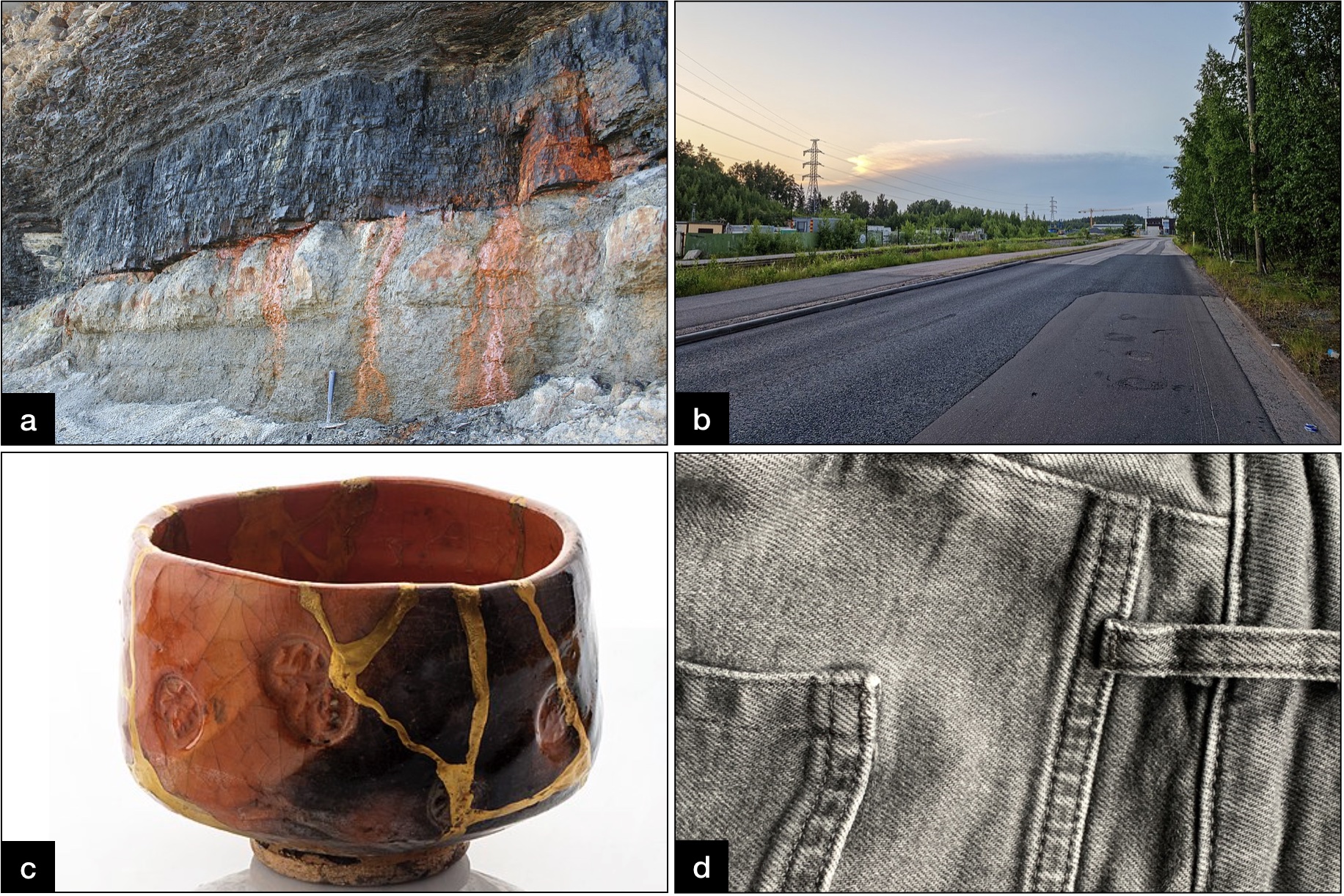}
  \caption{
  Examples of seams: a) naturally occurring coal seams within rock formations (photo by Michael Rygel, CC BY-SA 3.0, \url{https://commons.wikimedia.org/wiki/index/php?curid=12611040}); b) seams on a road caused by renewed asphalt concrete (photo by Coen, CC BY-SA 4.0,    \url{https://commons.wikimedia.org/w/index.php?curid=112442604}); c) a broken bowl restored with the Japanese art of kintsugi, leaving visible seams between the broken pieces (photo by Chiara Lorenzetti, CC BY-SA 4.0, \url{https://commons.wikimedia.org/w/index.php?curid=112661293}); d) seams holding together layers of fabric.}
  \Description{To exemplify the use and look of seams, this figure shows four examples of seams, both artifical seams, like in roadworks, crafts, and sewing, and naturally occursing seams, for example in layers of coal.}
\end{figure}
it.” Inman \& Ribes \cite{inman_beautiful_2019} suggest that seamful visualization may allow users to ask how data have been processed and how findings are produced. Addressing the messy relation between data and knowledge, Zer-Aviv \cite{Zer-Aviv_2014} argues: “[D]ata visualization should embrace the seamful approach that deliberately exposes the seams of the fallible human process of image making. One that acknowledges the image as an argument, as speech, as a part of visual language, to be debated with, questioned, talked back to or even visualized back to.” 

\section{Finding the seams in visualization}
While there hasn't been any dedicated literature that addresses seamful visualization beyond the considerations above, we can explore what seamful visualization could be, what it could express, and how. To do so, this paper takes three different approaches to review already existing occurrences of seams and seamfulness:
\begin{itemize}
\item exploring seams in realms outside of visualization and HCI;
\item considering visualization work already making use of seams;
\item tracing and contextualizing existing occurrences of seamfulness in visualization by drawing on qualities and themes of seamfulness from related work.
\end{itemize}
These approaches and the following exploration aim at a tentative take on seamful visualization rather than exhaustiveness.

\subsection{Seams}
To start exploring how seamful visualization and seams in critical visualization could look like we can refer to physical seams that we can find in contexts outside of visualization and HCI work.

Physical seams (see Fig. 1) are common in craftspersonship, like ceramics, sewing, mending, and kintsugi, in medical procedures, and in skilled trade works and construction, like welding, carpentry, and roadworks. While these examples highlight what seams, in a technical sense, can do, they don’t necessarily illustrate the deliberate use of seams. Rather, they illustrate how seams are utilized as a necessity, e.g., to fix rather than to remake, replace, or exchange something entirely, or stemming from a lack of alternatives. Most physical seams are artificial, but natural seams occur in geological phenomena, e.g., as coal seams or in crystals. If we think about the qualities of such seams, we can arrive at qualities of seamful visualization: seams can come into existence during the initial creation or during repairs, fixes, and maintenance. Seams can be used to fix cuts and tears, they can hold together different entities, layers, and where old and new meet; they can be places of fracture and they can be sealed, hidden, highlighted, or even decorative.

\subsection{Seams in visualization}
Seamful visualization has been underexplored, but we can find visualizations that—deliberately or not—make use of or represent seams in the more original meaning of the word (see Fig. 2). These are rather practical seams, emerging accidentally and being somewhat interrupting, or joining together two or more artifacts. 

For instance, seams emerge where physical maps are folded, in the middle of a world atlas page, or with (old) maps that don’t fit on one sheet \footnote{\url{https://maps.nls.uk/geo/explore/}}. Some maps of public transport systems display tariff zones by plotting lines on the map, representing the ‘seams’ between the different areas \footnote{E.g., \url{https://tfl.gov.uk/maps}, \url{https://www.bvg.de/en/connections/network- maps- and- routes}}. Visualizations juxtaposing two images side by side for a before and after comparison with a slider in the middle can also be considered as showing or making use of seams. German newspaper {\itshape Berliner Morgenpost} \footnote{\url{https://interaktiv.morgenpost.de/berlin-1953-2016/}} compares aerial imagery of Berlin from 1953 and 2016 and {\itshape Flightradar24} \footnote{\url{https://www.flightradar24.com/blog/then-and-now-visualizing-covid-19s-impact-on-air-traffic/}} visualizes the global air traffic before and during the pandemic. Albeit not strictly in the visualization realm, we can find seams in text, like in {\itshape Diagrammatic writing} \cite{drucker_diagrammatic_2013}, where Drucker experiments with the diagrammatic layout of text, refusing the linear presentation of text and strategically blending—or ‘stitching together’—paragraphs. Elsewhere, seams are wanted to be eliminated. For example, satellite or aerial imagery is usually joined together so that individual captures are invisible and one can only see ‘the whole’ \cite{haraway_situated_1988, kurgan_close_2013}.

\begin{figure}[h]
  \centering
  \includegraphics[width=0.9\linewidth]{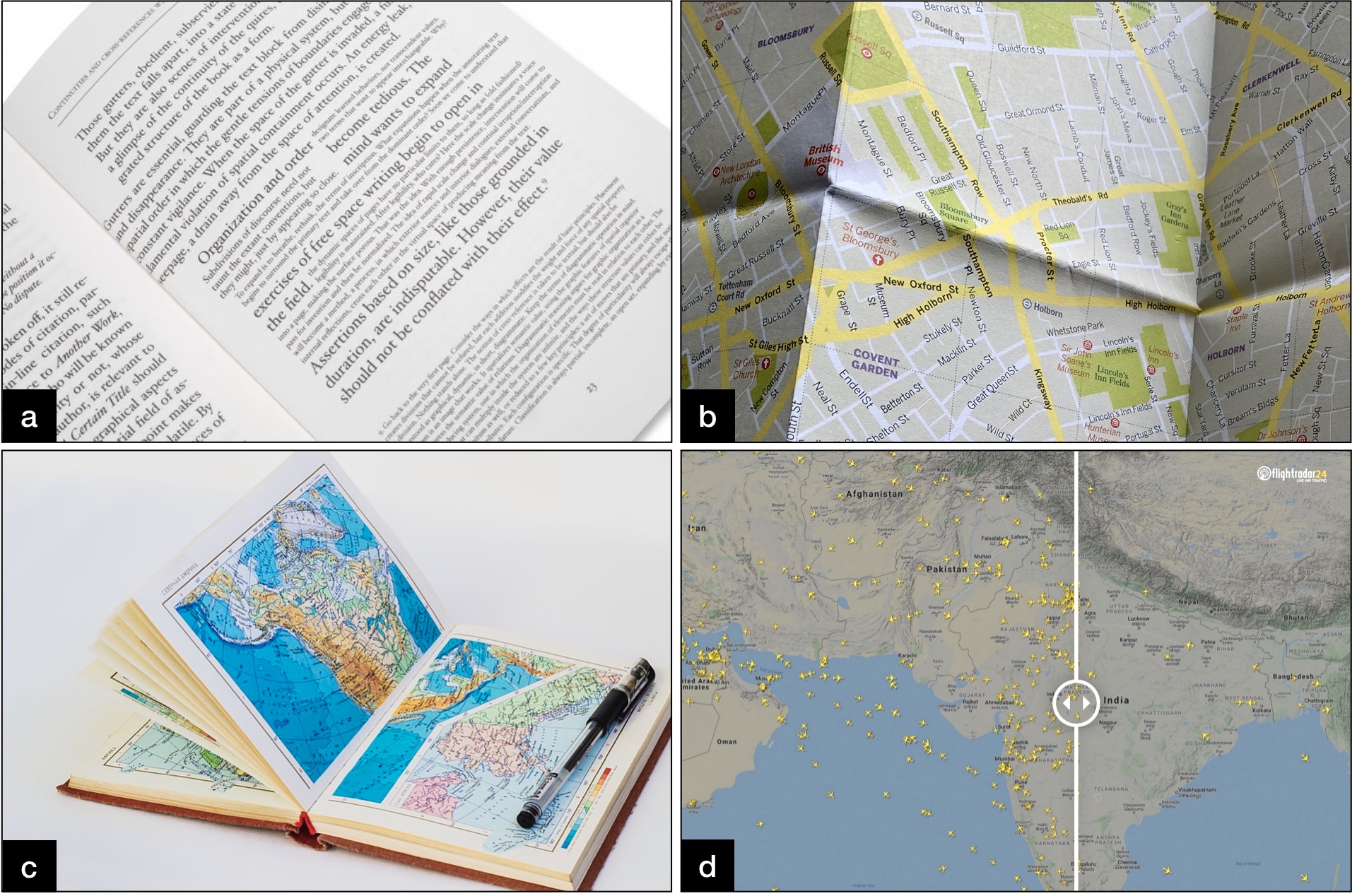}
  \caption{
  Examples of physical seams in visualization: a) a page of Diagrammatic writing showing paragraphs that are ‘stitched together’ in a non-linear layout (source: \url{https://www.onomatopee.net/exhibition/diagrammatic-writing/}, used with permission from onomatopee); b) a foldable map showing seams where it is un\slash folded; c) a seam in the middle of a world atlas page separates the map across two pages (photo by Anastasia Prisunko, CC BY-SA 4.0, \url{https://commons.wikimedia.org/w/index.php?curid=44941033}); d) a visualization representing air traffic before/after the pandemic divided by a slider (source: \url{https://www.flightradar24.com/blog/then-and-now-visualizing-covid-19s-impact-on-air-traffic/}, used with permission from Flightradar24.com).}
  \Description{This figure shows four examples of seams appearing in and around visualization. Three of these examples show seams that appear in printed visualization, e.g., in maps, Drucker's book on 'diagrammatic writing' and in a world atlas. The last example is a digital map that features an interactive slider in the middle of the map.}
\end{figure}

\subsection{Qualities of seamfulness}
By drawing together the qualities of seamfulness from the related work, we can trace how these, in a wider sense, apply to visualization or are already being addressed in theoretical and practice-oriented visualization work in both academia and industry. Based on the related work, seamfulness might allow for visualizations “to be debated with, questioned, talked back to or even visualized back to” \cite{Zer-Aviv_2014}, it may offer configurability or appropriation, or the “revelation of complexity, ambiguity or inconsistency” and consideration of internal operations and processes \cite{inman_beautiful_2019}. These qualities map onto the themes defined by Inman \& Ribes \cite{inman_beautiful_2019}: making visible what has previously been invisible, representing uncertainties, allowing for user appropriation, and offering time and interaction histories. For each of these affordances, we can consider existing visualization work in terms of how they express seamfulness.

\subsubsection{In\slash Visibility}
When thinking of seamful design as ‘making visible’, one may conclude that all visualizations are seamful, as they all make visible. While ‘making visible’ may apply to all visualizations and their ability to make visible patterns and structures in\slash of (large) data, there is always a choice to be made about what is made visible and what is left invisible, i.e., which features are highlighted or emphasized, and which are ignored or hidden. The examples in the following sections will illustrate how these choices are made and how this is driven by different approaches.

\subsubsection{Revelation of complexity}
Arguably, all visualizations represent complexity to some degree as most data are complex, but how this complexity is considered and addressed in visualization design differs. Some visualizations aim to specifically highlight the complexity of data and the represented phenomena through novel visual encodings and expressions. Surfacing this complexity within a visualization can be seen as making a visualization more seamful through surfacing the relations within the data and to external aspects. Where conventional visualizations don’t reflect on the qualities, situatedness, context, and relations of the represented data, the approaches below do.

In digital humanities, considering the complexity of data, material, and phenomena is central \cite{lamqaddam_introducing_2020, windhager_visualization_2018, windhager_exhibiting_2019}. For example, Drucker \cite{drucker_humanities_2011, drucker_non-representational_2017} argues that visualizations of humanist data\slash material “need to be more nuanced to show ambiguity and complexity” and proposes graphics that distort the seaming simplicity, certainty, and seamlessness that is usually prevalent in visualization. Drawing on Deleuze, Brüggemann et al. \cite{bruggemann_fold_2020} propose a critical framework for analysis and conceptualization of interactive visualizations that puts focus on the complexity of humanist data. They show examples of interactive folding operations that “offer insightful complications” \cite{bruggemann_fold_2020}, which visually resemble the appearance of seams.

\subsubsection{Uncertainty, ambiguity, and inconsistency}
Uncertainty visualization as a dedicated research area focuses on representing (statistically) uncertain, unclear, or ambiguous aspects of data and data analysis. Displaying statistical \cite{hullman_why_2019, kay_whenish_2016, kinkeldey_how_2014} or qualitative uncertainties \cite{bludau_relational_2019, panagiotidou_implicit_2021, windhager_uncertainty_2019} allows for a critical reading of data through surfacing some limitations of the data and offering visual cues that direct the interpretation of a visualization. This way, the otherwise seamless representation of data is interrupted and the gap between what the data show and how they should be interpreted is addressed.

McNutt et al. \cite{mcnutt_surfacing_2020} propose ‘visualization mirages’ to visually surface the failures and errors that can occur in the data analysis process. Instead of pretending that the analytics pipeline is a seamless and error-free process, this approach surfaces and opens the process’s seams. We can also consider visualizations that represent data rendered into sketch-like graphics \cite{wilber_roughviz_2022, wood_sketchy_2012}, which opposes the prevalent sleek graphics of visualizations and implies that the represented data may not be concise and that the visualization may be in flux, temporary, tentative, and unfinished.

\subsubsection{Revealing internal operations and processes}
While visualizations usually hide internal operations and processes, which adds to the seamless experience of data and their visualization, revealing parts of the internal operations and processes highlights the seams of the visualization process. It is somewhat common to disclose the data source or the involved organizations and people, but some work commits to a higher level of disclosure and transparency, aiming to contextualize and situate the visualization and to surface potential limitations of the data or method.

In their data storytelling project, Schwan et al.\footnote{\url{https://hannahschwan.de/inter...what-intersectionality!.html}} provide a disclosure layer in addition to the narrative layer, which offers additional information and context about the project, the positions and perspectives of the people involved, the design process and decisions, as well as critical reflections. In a similar vein, Burns \cite{burns_making_2021} explores disclosing metadata about visualizations, that is background information like the data source and data transformations, or how the visualization design came into existence. In a more systematic way, computational notebooks such as Observable, Jupyter Notebooks, or R Notebooks\footnote{\url{https://observablehq.com/}, \url{https://jupyter.org/}, \url{https://rmarkdown.rstudio.com/}} provide rich documentation that can include code, comments, narrative, and visualizations all in one place.

\subsubsection{Interaction histories}
History and provenance features in visualization generally aim at making a process more seamless, that is to make it easier to track changes or users, to collaboratively work on visualizations, or to bridge the gaps between different users, the technology, and its developers \cite{bavoil_vistrails_2005, cutler_trrack_2020, heer_graphical_2008, jankun-kelly_model_2002, shneiderman_eyes_1996}. 

Other work moves towards a more seamful approach of interaction histories and provenance by letting the audience reflect on their interaction with a visualization and by representing provenance of the visualized data, surfacing the seams between the interaction and the visualization, and within the visualization process. Vancisin et al. \cite{vancisin_externalizing_2020} propose provenance-driven visualization to represent the levels of transformations in historical documents and make visible the labor, interpretation, and curation of historical documents and data. Wall et al. \cite{wall_left_2021} visualize interaction traces by highlighting and summarizing the data points that a user has interacted with to raise “awareness of potential unconscious biases.” Feng et al. \cite{feng_hindsight_2016} contribute a similar system, offering cognitive support to change exploration behavior and “[encourage] people to visit more data and recall different insights after interaction.”

\subsubsection{Appropriation}
Adaptation, reuse, and (user) appropriation are not usually part of visualization designs, but are very much part of visualization tools, systems, or libraries. We can think of user appropriation in terms of visualizations and visualization tools that offer some level of personalization. On the author’s side of visualization, this personalization opens visualization processes at the usually hidden seams and makes it more accessible and thus contributes to the democratization of visualization. On the audience’s side, personalization offers a more personal, perhaps more empowering, and usable experience.

Representative of many visualizations that allow for personalization, {\itshape Show your stripes}\footnote{\url{https://showyourstripes.info/}} by Hawkins  lets viewers enter their geographical region so that the visualization represents the matching climate data. For visualization authors, GUI-based tools such as {\itshape datawrapper} and {\itshape RAWgraphs}\footnote{\url{https://www.datawrapper.de/}, \url{ https://rawgraphs.io/}} offer relatively easy ways of visualizing data and offer default settings as well as a range of reasonable design choices. Bostock’s {\itshape d3js} and {\itshape Observable}\footnote{\url{https://observablehq.com/}, \url{ https://d3js.org/}} project offer galleries of existing visualization examples that can be re-purposed by accessing and re-using the respective code.

\subsubsection{Visualizations that can be “debated with, questioned, talked back to or even visualized back to” \cite{Zer-Aviv_2014}}
Considering visualizations to offer a level of interactivity that includes feedback loops from the viewer to the visualization goes beyond the conventional interactive features that visualizations usually afford, such as filtering, zooming, selecting, undoing. By relying on some form of user input, some visualizations offer interactivity beyond these conventional terms and specifically surface and address the seams between the visualization and its audience.

Kauer et al. \cite{kauer_public_2021} found potential for participatory data visualizations that allow public audiences to add annotations, personal experiences, and stories, to encourage discussion of a visualization’s content. Similarly, ‘input visualizations’ \cite{huron_visualizations_2021} specifically require data input from an audience. Examples include public visualizations that aim at encouraging discourse in public spaces, e.g., through public displays \cite{claes_empowering_2017, coenen_citizen_2019, coenen_public_2021, koeman_urban_2017, koeman_what_2014, steinberger_vote_2014}, or data journalism articles that allow readers to draw on a plot to crowdsource assumptions, opinions, or knowledge so that it represents a form of discourse\footnote{Examples include \url{https://www.zeit.de/politik/deutschland/2021-09/bundestagswahl-parteien-ausrichtung-einschaetzung-konservatismus-progression-linke-rechte},\url{https://www.nzz.ch/storytelling/geografie-kenntnisse-wie-gut-koennen-sie-die- schweiz-aus-dem-gedaechtnis-zeichnen-ld.1306768}}.

\subsubsection{Summary}
The review of existing visualization work through the lens of seams and seamfulness shows what the conceptual lens of seamful visualization offers in terms of critically considering visualization design and evaluating what a visualization reveals and conceals. The examples discussed above are already making use of seamfulness one way or another and they do so by addressing seams on roughly three levels: seams between the visualization and the audience, seams between the data and what they represent, and seams within the processes of making, analyzing, interpreting, and interacting with data and visualizations. By making use of seams and seamfulness, the visualizations highlight the complexity, errors, uncertainties, or ambiguities of\slash in data, feature some level of disclosure and transparency regarding the visualization process, offer reflexive insights and engagement with the data’s limitations, afford personalization or re-use, or allow input from an audience to encourage discourse.

\section{The seamful design future of critical visualization}
The remaining sections conclude the paper by tying together the previous occurrences of seams and seamfulness and the (literal) meanings of seams, seamfulness, and seamlessness. By doing so, we can speculate how seamful visualization could look, what it could express, and how it can move us towards more expressive representations of data that account for the wider qualities and limitations of data.

\subsection{What could a seamful visualization express?}
To further develop and expand the idea of seamfulness in visualization we can consider what ‘seamless’ implies, how we can turn ‘seamless’ into ‘seamful’, and what physical seams imply and afford.

If thinking with seamlessness produces clear and clean visualization designs, thinking with seamfulness could produce messy—or at least messier—visualizations. We could reverse the seamlessness by integrating distractions, disruptions, discontinuities, inefficiencies, ineffectivenesses, complications, or inconsistencies; we could commit to flawed, imperfect, or blemished designs, or find value in unclean and unclear layouts.

Like physical seams, we might use seams to hold together, make visible, or compare different layers, entities, or views in visualization, to intentionally highlight certain aspects, or to guide the reading and interpretation of a visualization. In theory, we could surface seams of one or more of the stages and contexts of the data--visualization process: data creation, code, data (set, setting, analysis, wrangling), visualization (design, interaction, interpretation). This could mean highlighting code and what it does or showing errors in the code. It might mean strategically highlighting irregularities and ambiguities in data; making considerable the ‘raw’ data, the provenance of the data, how they were cleaned, processed; highlighting errors, uncertainties, ambiguities, discrepancies, and parts of the data that have been rejected\slash discarded; emphasizing errors, friction, and where data don’t add up. Surfacing seams may involve adding an alternative layer or contexts; including reference points from the real-world phenomenon; revealing the evolution of the visualization, which designs have been discarded and why; displaying different base maps, perspectives, or scales; showing different states of the data, e.g., before and after ‘cleaning’, old and new data, or different data altogether.

\subsection{Principles for seamful critical visualization}
To further engage with seamful visualization, both in practice and future research, we can consider the following tentative principles.

\textbf{Finding the seams.} Tracing the limitations of the data and the qualities, i.e., seams, of the data–phenomenon relationship enables us to represent data more sincerely in a visualization. As some—perhaps most—of the limitations and qualities of data aren’t apparent or traceable from a distance, we can engage with data studies \cite{kitchin_data_2021, kitchin_towards_2014, Richterich_2018}, close readings \cite{loukissas_place_2016}, and other (qualitative) research methods \cite{masson_data_2020} to understand how data do and do not represent what they represent.

\textbf{Encouraging critical engagements.} In quest of critical visualization, seamfulness should encourage critical engagements with data and their visualization. This might mean to commit to designs that actively encourage the audience to pose questions about the data, for example, by making considerable data’s limitations and qualities, by drawing attention to the less clear and messier aspects of data, data practices, and data–phenomenon relationships, or by communicating what the data do not show.

\textbf{Resisting the seamlessness.} To explore the value of seamful visualization and to find richer, more expressive approaches to representing data, we should question seamlessness and attend to seamfulness. This might mean to experiment with strategically concealing and revealing aspects of data, opening otherwise hidden seams of the involved processes and practices, or intentionally disrupting\slash interrupting the (seeming) seamlessness through seamful design strategies.

\textbf{Performing the balancing act.} Seamful visualization might mean to make a visualization more seamful, rather than to make it fully seamful. This suggests a balancing act between simplicity and complexity, seamlessness and seamfulness: while an entirely seamless visualization hides complexity and accepts oversimplification, an entirely seamful visualization wouldn’t be productive, readable, or usable, and would negate the cognitive and perceptual capacities of visualization. To make use of seamfulness in visualization, we need to explore how the blend and the capacities of seamful and seamless elements and features depend on the data, the purpose, and the context of a visualization. 

\begin{acks}
I thank the reviewers for their valuable feedback, Greg McInerny and João Porto de Albuquerque for supervising this work, the organizers and participants of the Information+ conference 2021 for the productive discussion on seamlessness and seamfulness in visualization, and Mushon Zer-Aviv for drawing my attention to the concept of sketch-like renderings of visualizations. This work is supported by the Centre for Doctoral Training in Urban Science and Progress, funded by the UK Engineering and Physical Science Research Council under grant no. EP/L016400/1.
\end{acks}

\bibliographystyle{ACM-Reference-Format}

\end{document}